\documentclass[prl,showpacs,preprintnumbers,twocolumn]{revtex4}
\usepackage{amssymb}
\usepackage{color}
\usepackage{graphicx}

\begin{document}

\title[Spin Wave]{Spin-wave mediated quantum corrections to the conductivity
in thin ferromagnetic gadolinium films}
\author{R. Misra, A.F. Hebard}
\email[Corresponding author:~]{afh@phys.ufl.edu}
\author{K.A. Muttalib}
\affiliation{Department of Physics, University of Florida, Gainesville FL 32611}
\author{P. W\"{o}lfle}
\affiliation{ITKM, Universit\"{a}t Karlsruhe, D-76128 Karlsruhe, Germany}
\affiliation{INT, Forschungszentrum Karlsruhe, D-76021 Karlsruhe, Germany}
\author{}

\begin{abstract}
We present a study of quantum corrections to the conductivity of thin
ferromagnetic gadolinium films. \textit{In situ} magneto-transport
measurements were performed on a series of thin films with thickness $d <
135 A$. For sheet resistances $R_0 < 4011 \Omega$ and temperatures $T<30 K$,
we observe a \textit{linear} temperature dependence of the conductivity in
addition to the logarithmic temperature dependence expected from well-known
quantum corrections in two dimensions. We show that such a linear $T$%
-dependence can arise from a spin-wave mediated Altshuler-Aronov type
correction.
\end{abstract}

\maketitle
\date{May 2008}

Recent studies of the anomalous Hall effect in thin Fe films \cite{mitra}
have provided strong evidence for weak localization (WL) effects in
disordered ferromagnetic films. Since WL effects are cut off by various
temperature independent phase breaking scatterings and especially by the
magnetic field inside the ferromagnet, one needs a sufficiently large
temperature dependent phase relaxation rate $1/\tau_{\phi}$ to have an
experimentally accessible disorder and temperature interval where such
effects can be observed. While the contribution to $%
1/\tau_{\phi}$ in ferromagnetic films from electron-electron interactions is
small, a much larger contribution is obtained from scattering off spin-waves 
\cite{tatara,plihal}, such that the characteristic logarithmic temperature
dependence of the conductivity due to WL effect was observed in
polycrystalline Fe films within a range of temperature $5K <T< 20K$, for
sheet resistances $R_0< 3k\Omega$.

Given the importance of spin-waves in Fe films, one expects to see an even
larger effect in other ferromagnetic films with larger magnetic moments. In
particular, quantum corrections to the conductivity due to scattering off
spin-waves should be observable, just like the quantum corrections due to
electron-electron Coulomb interactions, if the exchange coupling is large
enough and the spin-wave gap is smaller than the temperature. It turns out
that an excellent candidate is gadolinium (Gd), with a spin-wave gap of
about $30~mK$  and a Curie temperature of $293K$ \cite{mukho, coqblin}.

We have carried out systematic \textit{in situ} magnetotransport
measurements on a series of Gd films with varying thicknesses ($35%
\mathring{A}<d<135 \mathring{A}$) having sheet resistances $R_0$ ranging
from $428 \Omega$ to $4011 \Omega$. For temperatures $5K< T < 30 K$, we
observe simultaneous presence of two types of quantum corrections to the
Drude conductivity: one has the expected logarithmic temperature dependence
that is a hallmark of quantum corrections in two-dimensions \cite{lee}; the
other has a heretofore-unobserved approximately linear temperature dependence,
which we
attribute to a spin-wave mediated Altshuler-Aronov type correction to the
conductivity. We calculate this spin-wave contribution within a standard
diagrammatic perturbation theory and show that the results agree with the
experimentally observed temperature dependence.

A series of ultrathin films of Gd in the Hall bar geometry was grown by r.f.
magnetron sputtering through a shadow mask onto sapphire substrates held at
a temperature of 250~K. The current and voltage leads of the deposited
sample overlapped with predeposited palladium contacts, thus allowing
reliable electrical connection for \textit{in situ} measurements of the
electrical properties. The experiments were performed in a specialized
apparatus in which the sample can be transferred without exposure to air
from the high vacuum deposition chamber to a 7\,T magnet located in a low
temperature cryostat. To parameterize the amount of disorder in a given film%
\cite{mitra}, we use sheet resistances $R_{0}\equiv R_{xx}(T=5K)$ where $%
R_{xx}$ is the longitudinal resistance. $R_0$ spans the range from $%
428\Omega $ (135 {\AA } thick) to $4011\Omega $ ( 35 {\AA } thick).
Longitudinal and Hall resistances were measured using standard four-probe
lock-in techniques for low resistance samples ($R_0 \leq 2613\Omega$) and dc
techniques for higher resistance samples($R_0 > 2613\Omega$).

The inset of Fig.~1 shows the temperature dependence of the longitudinal conductivity
$\sigma_{xx}$ for a series of thin Gd films with varying sheet resistances
$R_0$ listed in the legend.  
\begin{figure}[tbp]
\begin{center}
\includegraphics[angle=0, width=0.35\textheight]{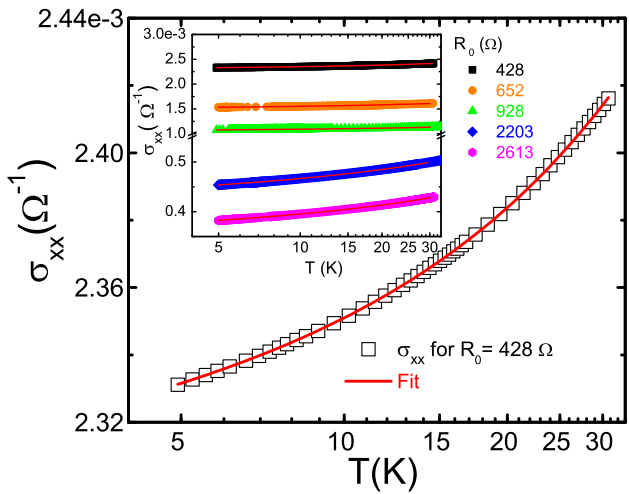}
\end{center}
\caption{Temperature dependence of $\protect\sigma_{xx}$ for a series of Gd
thin films (inset) with various sheet resistances $R_0$ listed in the legend.
Note that the temperature scales
are logarithmic. The fits for each curve are obtained using Eq.~(\protect\ref{fit}), with
fitting parameters listed in Table I. The main panel in an expanded view shows the quality of fit
for the $R_0 = 428 \Omega$ data.}
\label{fig-T}
\end{figure}
A fitting function of the form 
\begin{equation}  \label{fit}
\sigma_{xx}/ L_{00} = P_1+P_2\ln (T /T_0)+P_3\left( T / T_0 \right)^p,
\end{equation}
where $L_{00}=e^2/\pi h$ and $T_0=5K$ is a reference temperature, has been
used to fit each of the curves. An expanded view of a typical fit (solid red line) is
shown in the main panel for the $R_0 = 428 \Omega$ data (open squares).
The fitting parameters for all of the curves are shown in Table I with
respective errors in parentheses.

\begin{table}[b!]
\begin{center}
\begin{tabular}{lcccccccc}
\hline
$R_{0}(\Omega)$ & $P_1$ &  & $P_2$ &  & $P_3$ &  & $p$ &  \\ \hline
$428$ & $187.88(1)$ &  & $0.79(1)$ &  & $0.97(1)$ &  & $1.039(6)$ &  \\
$652$ & $123.24(1)$ &  & $0.67(1)$ &  & $1.03(1)$ &  & $0.976(5)$ &  \\
$928$ & $86.47(1)$ &  & $0.71(1)$ &  & $0.72(2)$ &  & $0.934(8)$ &  \\
$2203$ & $36.10(2)$ &  & $0.75(1)$ &  & $0.65(2)$ &  & $0.84(1)$ &  \\
$2613$ & $30.25(1)$ &  & $0.70(1)$ &  & $0.72(1)$ &  & $0.818(4)$ &  \\ \hline
\end{tabular}%
\end{center}
\caption{Fitting parameters defined in Eq.~(\protect\ref{fit}) used in Fig.~%
\protect\ref{fig-T}}
\label{table1}
\end{table}

There are several important points about the fits that are worth emphasizing.
First, note that the temperature scale is logarithmic, and it is clear that
a $\ln T$ alone can not fit the data. Second, The power $p$ is close to $1$
for $R_0 < 3k\Omega$, but then changes significantly when $R_0>4 k\Omega$.
Third, the coefficient $P_3$ is roughly independent of disorder. We will
show that the unusual temperature dependence is consistent with a sum of
contributions from well-known quantum corrections in two-dimensions and a
novel spin-wave mediated correction analogous to the Altshuler-Aronov
electron-electron contribution in disordered systems \cite{altschuler}. While the
Altshuler-Aronov contribution gives rise to a logarithmic temperature
dependence in two dimensions, we show that the spin-wave mediated
contribution can be linear in temperature within certain ranges of the
parameters, consistent with the experiments. The theory ceases to be valid for
large disorder ($R_0>4 k\Omega$), where the temperature dependence is no
longer linear.

To make comparisons with earlier studies on Fe films \cite{mitra,by}, we
have also measured the Anomalous Hall (AH) resistances of the Gd films at 7
Tesla magnetic field. Following Ref.~[\onlinecite{by}], we define normalized
relative changes 
\begin{equation}  \label{nrc}
\Delta^{N}(Q_{ij})\equiv (1/L_{00}R_{0})(\delta Q_{ij}/Q_{ij})
\end{equation}
with respect to a reference temperature $T_{0}= 5K <T$, where $\delta
Q_{ij}=Q_{ij}(T)-Q_{ij}(T_{0})$ and $Q_{ij}$ refers to either resistances $%
R_{xx},R_{xy}$ or conductivity $\sigma _{xx},\sigma _{xy}$.

Figure 2 shows $\Delta^N(\sigma_{xy})$, together with $\Delta^N(R_{xx})$ and 
$\Delta^N(R_{xy})$ for comparison, for two different sheet resistances. 
\begin{figure}[bp]
\begin{center}
\includegraphics[angle=0, width=0.35\textheight]{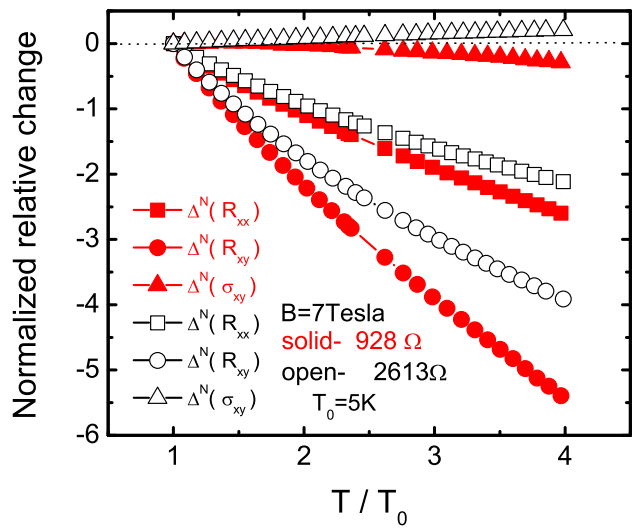}
\end{center}
\caption{Normalized relative changes $\Delta^N(\protect\sigma_{xy})$,
defined in Eq.~(\protect\ref{nrc}), for two different sheet resistances. For
comparison we also show $\Delta^N(R_{xx})$ and $\Delta^N(R_{xy})$. }
\label{fig-nrc}
\end{figure}
We find that within the range of disorder, $\Delta^N\sigma_{xy}\approx 0$
for $5K<T<20K$. As shown in Ref. [\onlinecite{mw}], the interaction
correction to the AH conductivity is exactly zero, due to symmetry reasons.
This is true for both repulsive Coulomb interaction and the attractive
spin-wave mediated interaction. However, the WL correction to the AH
conductivity need not be zero. In fact, the total WL contribution is given
by 
\begin{equation}
\Delta^N\sigma^{WL}_{xy}=\frac{1}{1+r_{xy}}\ln\frac{T}{T_0}; \;\;\;
r_{xy}\equiv \frac{\sigma_{xy}^{sj}}{\sigma_{xy}^{ss}}
\end{equation}
where $r_{xy}$ is the ratio of two different mechanisms contributing to the
AH conductivity, namely the side-jump \cite{berger} $\sigma_{xy}^{sj}$ and
the skew scattering \cite{smit} $\sigma_{xy}^{ss}$. Thus the ratio $r_{xy}$
is a non-universal quantity. It turns out that for Fe films, while $r_{xy}
>> 1$ for films deposited on glass, the opposite $r_{xy} << 1$ is true for
films deposited on sapphire \cite{mitra} or on antimony \cite{by}. Our
current results $\Delta^N\sigma_{xy}\approx 0$ for Gd deposited on sapphire
agree with those of Fe films deposited on the same substrate. As for the
longitudinal part, the coefficient $A_R$ defined by 
\begin{equation}
\Delta^N\sigma_{xx}=A_R\ln (T/T_0)
\end{equation}
is given as $A_R=A_R^{WL}+A_R^I=1+h_{xx}$ where the first term ($A_R^{WL}=1$%
) is due to WL and the second term ($A_R^I=h_{xx}$) is the exchange plus
Hartree interaction contribution, with $h_{xx}=(1-3\tilde{F})/4$ where $%
\tilde{F}$ is the Hartree term. It has been argued that at least for Fe
films, the total interaction correction $h_{xx}$ is very small due to a near
cancellation of the exchange and Hartree terms, which is expected due to
strong screening. This results in $A_R \approx 1$ for Fe films. Table I
shows that $A_R \approx 0.75$ for our Gd films. This suggests that $h_{xx}$
may actually be negative, due to an even larger Hartree contribution. Since
both $r_{xy}$ and $h_{xx}$ are non-universal quantities, we only note that a
large $r_{xy}$ (similar to Fe films on saphhire) and a small negative $h_{xx}
$ (large Hartree term, again similar to Fe films) are consistent with the
current experimental observations (on Gd films on sapphire).

To understand the linear $T$-dependence of the longitudinal conductivity, we
evaluate the spin wave contributions within the standard diagrammatic
perturbation theory. The film is described as a quasi-two-dimensional system
of conduction electrons with Fermi energies $\epsilon _{F\sigma }$ depending
on the spin index $\sigma =\uparrow ,\downarrow $. We model the total
impurity potential as a sum over identical single impurity potentials $V(%
\mathbf{r-R}_{j})$ at random positions $\mathbf{R}_{j}$. The Hamiltonian is
given by 
\begin{eqnarray}
H &=& {\textstyle\sum\limits_{\mathbf{k}\sigma }}(\epsilon _{\mathbf{k} }- 
\frac{1}{2}\sigma B)c_{\mathbf{k}\sigma }^{+}c_{\mathbf{k}\sigma }\cr &+& {%
\textstyle\sum\limits_{\mathbf{k},\mathbf{k}\prime \sigma, \mathbf{j}}}V(%
\mathbf{k-k}^{\prime })e^{i(\mathbf{k-k}^{\prime })\cdot \mathbf{R}_{j}}c_{%
\mathbf{k\prime\sigma }}^{+}c_{\mathbf{k}\sigma } \cr &+& \sum_q\omega_q
a_q^+a_q + J\sum_{q,k}[a_q^+c^+_{k+q\downarrow}c_{k\uparrow}+h.c.]
\end{eqnarray}
where $c_k$,$\;c_k^+$ are electron field operators and $a_q$,$\;a_q^+$ are
spin-wave operators and $J$ is the effective spin exchange interaction. The
spin wave is characterized by $\omega_q=\Delta_g+Aq^2$, where $%
\Delta_g\approx\mu_BB_{ext} \approx 1K/Tesla$ is the spin-wave gap and $%
A\approx J/k_F^2$ is the spin stiffness. Later we will drop $\Delta_g$ as $%
\Delta_g<T$. The exchange splitting $B\approx Jk_F^2$ is large, but $%
B/\epsilon_F << 1$, where we have used the values $B = 700 meV$ at 20K \cite{bode}
and $\epsilon_F = 3.4 eV$ \cite{coqblin}. The spin-wave propagator is 
\begin{equation}
S_{\uparrow\downarrow}^{SW}(q,\omega_n)= 1/(i\omega_n-a\omega_q) =
[S_{\downarrow\uparrow}^{SW}]^*
\end{equation}
where $\omega_n=2\pi nT$ is the bosonic Matsubara frequency, $a=1-i\gamma/2$
and $\gamma$ is a phenomenological damping constant. The spin-wave mediated
effective interaction is given by $V_{sw}(q,\omega_n)=nJ^2[S_{\uparrow%
\downarrow}^{SW}(q,\omega_n) + S_{\downarrow\uparrow}^{SW}(q,\omega_n)]$
which is attractive. Here $n$ is the density of conduction electrons.

For the quantum corrections to the conductivity, the dominant contributions
from spin-wave interactions come from the diagrams with the most number of
diffusons, analogous to those relevant for the Coulomb interactions. We
therefore first evaluate the diffuson propagator $\Gamma^{\uparrow%
\downarrow}(q,\omega)$ in the presence of a large exchange splitting, and
obtain
\begin{eqnarray}
\Gamma^{\uparrow\downarrow}(q,\omega_n)=1/[(2\pi N_0\tau\hat{\tau}) 
(\omega_n-iB+\hat{D}q^2)]
\end{eqnarray}
where $N_0=m/2\pi$ is the density of states at the Fermi surface with $m$
being the electron mass, $\tau$ is the scattering time, and we have defined 
\begin{eqnarray}
\hat{D}\equiv D \left( \hat{\tau}/\tau \right)^2; \;\;\; 1/\hat{%
\tau}\equiv 1/{\tau}+\omega_n-iB
\end{eqnarray}
where $D=(1/2)v_F^2\tau$ is the diffusion coefficient. The corresponding $%
\Gamma^{\downarrow\uparrow}(q,\omega_n)$ is obtained by replacing $B$ by $-B$
everywhere. 
\begin{figure}[tbp]
\begin{center}
\includegraphics[angle=0, width=0.3\textheight]{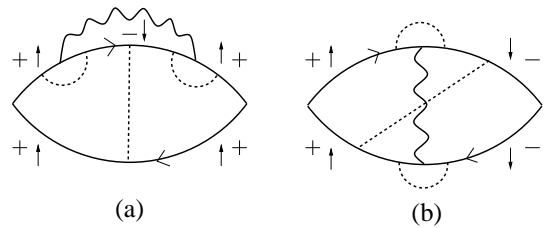}
\end{center}
\caption{Spin-wave contributions to the longitudinal conductivity. Solid
lines are impurity averaged Green's functions, broken lines are diffusons,
wavy line represents the effective spin-wave mediated interactions. There
are two diagrams of type (a) and four diagrams of type (b)}
\label{fig-sw}
\end{figure}

From the diagrams shown in Figure 3, the spin wave contribution to the
longitudinal conductivity is given, e.g. for diagram 3a, by 
\begin{eqnarray}
\delta \sigma _{xx \uparrow\downarrow}^{SW} &=&
T\sum_{\omega_n}\int_{0}^{q_c} \frac{dq^2}{(2\pi)^2}q^2v_F^2(2\pi N_0\tau%
\hat{\tau}^2)^2(2\pi N_0\tau)^2 \cr &\times& \left[\Gamma^{\uparrow%
\downarrow}(q,\omega_n)\right]^3 S_{\uparrow\downarrow}^{SW}(q,\omega_n)
\end{eqnarray}
and similarly for the $\downarrow\uparrow$ case. Here the upper cut-off in
the $q$-integral is given by $q_c = |1/v_F\hat{\tau}|$. The total spin-wave
contribution can then be written as 
\begin{eqnarray}
\frac{\delta \sigma _{xx}^{SW}}{ L_{00}}\approx \frac{4N_0J}{1+\gamma^2/4}%
\frac{nJ}{B} \frac{\epsilon_F}{B}(\epsilon_F \tau) \frac{T}{Ak_F^2}
\end{eqnarray}
where we have assumed that $B\tau >> 1$.

Using $n=k_F^2/4\pi$, we estimate $\delta \sigma _{xx}/L_{00} \approx (\frac{%
Jk_F^2}{2\pi B})^2(\epsilon_F\tau) \frac{T}{Ak_F^2}$ for small damping. With
an estimate of $\epsilon_F\tau\sim 10$ and $(Jk_F^2)/ (2\pi B)\sim 1$, the
observed magnitude of the constant $P_3\sim 1$ is consistent if $%
T_0/Ak_F^2\sim1/10$, which is quite reasonable. On the other hand, the
disorder dependence of the linear $T$ contribution is given by $P_3\sim
\epsilon_F\tau$, which decreases with increasing disorder. However, as
observed experimentally, $P_3$ appears to decrease only weakly with disorder up to
a sheet resistance $R_0=2613 \Omega$.
The variation in sheet resistance in the regime considered here may be mainly 
due to the change in film thickness, while the disorder strength varies only weakly.
We note
that while the linear $T$ behavior is observed to be quite robust for weak
disorder, it cannot explain the data for $R_0>4k\Omega$ (not shown in this 
paper). We expect that for higher sheet resistances the system will undergo 
an Anderson localization transition from the pseudo-metallic phase 
(localization length longer than the sample dimensions) to a truely 
localized phase. The study of that regime is the subject of a forthcoming 
publication.

In conclusion, we have studied charge transport in ultrathin films of Gd
grown using \textit{in-situ} techniques which exclude in particular unwanted oxidation or
contamination. In addition to the logarithmic temperature dependence
expected from the weak localization effects in the longitudinal conductivity
as previously seen in Fe films, we observe an additional contribution to the
conductivity that has an approximately linear $T$-dependence
for sheet resistance \textit{\ }$R_{0}<4k\Omega$ and
temperatures $5K<T<30K$. We interpret this novel feature in terms of
contributions from scattering off spin-waves which are known to be important
in ferromagnetic films. We find from our calculations that the interaction of 
the electrons with spin waves of the ordered ferromagnet gives rise to a 
much larger contribution than the usual one generated by the Coulomb 
interaction. The temperature dependence is governed by the singular (in the 
limit $\omega, q  \rightarrow  0$) spin wave propagator. The dressing by diffuson 
lines changes the prefactor, but leaves the temperature dependence unchanged.
This is, to our knowledge, the first time that this type of quantum 
correction has been seen and explained. 

This work has been supported by the NSF under Grant No. 0704240 (AFH), and
by the DFG-Center for Functional Nanostructures (KAM, PW).

\end{document}